\documentclass[english]{revtex4}
\usepackage[T1]{fontenc}
\usepackage[latin9]{inputenc}

\makeatletter

\usepackage{babel}
\makeatother

\begin{document}

\title{On the Thermodynamical Relation between Rotating Charged BTZ Black
Holes and Effective String Theory}

\author{Alexis Larrañaga}

\email{eduardalexis@gmail.com}

\affiliation{National University of Colombia}

\affiliation{National Astronomical Observatory (OAN)}

\begin{abstract}
In this paper we study the first law of thermodynamics for the (2+1)
dimensional rotating charged BTZ black hole considering a pair of
thermodinamical systems constructed with the two horizons of this
solution. We show that these two systems are similar to the right
and left movers of string theory and that the temperature associated
with the black hole is the harmonic mean of the temperatures associated
with these two systems. 
\end{abstract}
\maketitle

\section{Introduction}

Bekenstein and Hawking showed that black holes have non-zero entropy
and that they emit a thermal radiation that is proportional to its
surface gravity at the horizon. When the black hole has other properties
as angular momentum $\mathbf{J}$ and electric charge $Q$, these
quantities are related with the mass through the identity

\begin{equation}
dM=TdS+\Omega dJ+\Phi dQ,\end{equation}

where $\Omega=\frac{\partial M}{\partial J}$ is the angular velocity
and $\Phi=\frac{\partial M}{\partial Q}$ is the electric potential.
This relation is called \emph{the first law of black hole thermodynamics}\citet{hawking,bekenstein}.
When the black hole has two horizons, as in the case of Kerr-Newman
solution, it is known that it is possible to associate a thermodynamical
first law with each of them. The outer horizon is related with the
Hawking radiation while the inner horizon is related with the absortion
proccess \citet{Zhao,Wu,Wu1}.\\

In this paper we will apply the method described by Wu \citet{wu2}
to describe the thermodynamics of the rotating charged BTZ black hole
in (2+1) dimensions and relate it with the effective string theory
and D-brane description of black holes. It is well known that a characteristic
of effective strings or D-branes is that their spectrum can be expressed
as right and left moving excitations. The entropy of ther system is
given as the sum of the two contributions,

\begin{equation}
S=S_{R}+S_{L},\end{equation}
while the temperature associated with the system is the harmonic mean
of the temperatures of the modes,

\begin{equation}
\frac{2}{T_{H}}=\frac{1}{T_{R}}+\frac{1}{T_{L}}.\end{equation}

Therefore, we will define two thermodynamical systems as the sum and
the difference of the two horizons associated with the rotating charged
BTZ black hole. These systems resemble the R and L moving modes of
string theory and will provide a way to show how the Hawking temperature
$T_{H}$ associated with the BTZ black hole can be interpreted as
the harmonic mean of the temperature of the R and L parts. We will
also identify the electric potential and the angular velocity associated
with each of the systems and their relation with the black hole properties.

\section{The Rotating Charged BTZ Black Hole}

The standard rotating charged BTZ black hole \citet{martinez,banados,banados2}
is a solution of $\left(2+1\right)$ dimensional gravity with a negative
cosmological constant $\Lambda=-\frac{1}{l^{2}}$. Its line element
can be written as

\begin{equation}
ds^{2}=-\Delta dt^{2}+\frac{dr^{2}}{\Delta}+r^{2}d\varphi^{2},\end{equation}

where the lapse function is

\begin{equation}
\Delta=-M+\frac{r^{2}}{l^{2}}+\frac{J^{2}}{4r^{2}}-\frac{Q^{2}}{2}\ln\left(\frac{r}{l}\right).\end{equation}

This solution has two horizons given by the condition $\Delta=0$,

\begin{equation}
M=\frac{r_{\pm}^{2}}{l^{2}}+\frac{J^{2}}{4r_{\pm}^{2}}-\frac{Q^{2}}{2}\ln\left(\frac{r_{\pm}}{l}\right).\label{eq:radii}\end{equation}

The Bekenstein-Hawking entropy associated with the black hole is twice
the perimeter of the outer horizon,

\begin{equation}
S=4\pi r_{+},\end{equation}
and therefore, the mass can be written in terms of this entropy as

\begin{equation}
M=\frac{S^{2}}{\left(4\pi\right)^{2}l^{2}}+\frac{\left(4\pi\right)^{2}J^{2}}{4S^{2}}-\frac{Q^{2}}{2}\ln\left(\frac{S}{4\pi l}\right).\label{eq:massformula}\end{equation}

The Bekenstein-Smarr integral mass formula is \citet{larr,wang}

\begin{equation}
M=\frac{1}{2}TS+\Omega J+\frac{1}{2}\Phi Q+\frac{1}{4}Q^{2},\end{equation}

or

\begin{equation}
M=\kappa\mathcal{P}+\Omega J+\frac{1}{2}\Phi Q+\frac{1}{4}Q^{2},\label{eq:generalIntFirstLaw}\end{equation}

where $\mathcal{P}=\frac{P}{2\pi}$ is the {}``reduced'' perimeter
and $\kappa$ is the surface gravity. Thus, the Hawking-Bekenstein
entropy can be written as

\begin{equation}
S=4\pi\mathcal{P},\end{equation}
and the mass (\ref{eq:massformula}) is given by

\begin{equation}
M=\frac{\mathcal{P}^{2}}{l^{2}}+\frac{J^{2}}{4\mathcal{P}^{2}}-\frac{Q^{2}}{2}\ln\left(\frac{\mathcal{P}}{l}\right).\end{equation}

Finally, the differential form of the first law for this black hole
takes the form \citet{Akbar}

\begin{equation}
dM=2\kappa d\mathcal{P}+\Omega dJ+\Phi dQ.\label{eq:generalDiffFirstLaw}\end{equation}

As discussed before \citet{Wu,larr2}, we can associate a thermodynamics
to both outer and inner horizons. The four laws associated with theser
horizons describe the Hawking radiation process as well as the absortion
process. Therefore, the integral and differential mass formulae can
be written for the two horizons,

\begin{eqnarray}
M & = & \frac{\mathcal{P_{\pm}}^{2}}{l^{2}}+\frac{J^{2}}{4\mathcal{P_{\pm}}^{2}}-\frac{Q^{2}}{2}\ln\left(\frac{\mathcal{P}_{\pm}}{l}\right)\label{eq:massequations}\\
dM & = & 2\kappa_{\pm}d\mathcal{P}_{\pm}+\Omega_{\pm}dJ+\Phi_{\pm}dQ.\end{eqnarray}
From these relations, is easy to see that the surface gravity and
electrostatic potential at the two horizons are

\begin{eqnarray}
\kappa_{\pm} & = & \frac{1}{2}\left.\frac{\partial M}{\partial\mathcal{P}_{\pm}}\right|_{Q}=\frac{\mathcal{P_{\pm}}}{l^{2}}-\frac{J^{2}}{4\mathcal{P_{\pm}}^{3}}-\frac{Q^{2}}{4P_{\pm}}=\frac{r_{\pm}}{l^{2}}-\frac{J^{2}}{4r_{\pm}^{3}}-\frac{Q^{2}}{4r_{\pm}}\\
\Phi_{\pm} & = & \left.\frac{\partial M}{\partial Q}\right|_{\mathcal{P}_{\pm}}=-Q\ln\left(\frac{\mathcal{P}_{\pm}}{l}\right)=-Q\ln\left(\frac{r_{\pm}}{l}\right)\\
\Omega_{\pm} & = & \left.\frac{\partial M}{\partial J}\right|_{\mathcal{P}_{\pm}}=\frac{J}{2\mathcal{P_{\pm}}^{2}}=\frac{J}{2r_{\pm}^{2}},\end{eqnarray}

while the entropy and temperature associated with each horizon are

\begin{eqnarray}
S_{\pm} & = & 4\pi\mathcal{P}_{\pm}\\
T_{\pm} & = & \frac{\kappa_{\pm}}{2\pi}.\end{eqnarray}

From equation (\ref{eq:massequations}) we can obtain to important
relations,

\begin{equation}
M=\frac{\mathcal{P}_{+}^{2}+\mathcal{P}_{-}^{2}}{2l^{2}}+\frac{J^{2}}{8}\left[\frac{1}{\mathcal{P}_{+}^{2}}+\frac{1}{\mathcal{P}_{-}^{2}}\right]-\frac{Q^{2}}{4}\ln\left(\frac{\mathcal{P}_{+}\mathcal{P}_{-}}{l^{2}}\right)\label{eq:sum}\end{equation}

\begin{equation}
\frac{\mathcal{P}_{+}^{2}-\mathcal{P}_{-}^{2}}{l^{2}}+\frac{J^{2}}{4}\left[\frac{1}{\mathcal{P}_{+}^{2}}-\frac{1}{\mathcal{P}_{-}^{2}}\right]=\frac{Q^{2}}{2}\ln\left(\frac{\mathcal{P}_{+}}{\mathcal{P}_{-}}\right)\label{eq:diff}\end{equation}

Now, using the inner and outer horizons we will define two independient
thermodynamical systems. Following Wu \citet{wu2}, the R-system will
have a reduced perimeter corrspondient to the sum of the inner and
outer perimeters while L-system corresponds to the difference of these
perimeters,

\begin{eqnarray}
\mathcal{P}_{R} & = & \mathcal{P}_{+}+\mathcal{P}_{-}\\
\mathcal{P}_{L} & = & \mathcal{P}_{+}-\mathcal{P}_{-}.\end{eqnarray}

It is important to note that each of these systems carry three hairs
$\left(M,J,Q\right)$. Now, we will obtain the thermodynamical relations
for these systems and then, we will relate them with the thermodynamics
of the charged BTZ black hole and its Hawking temperature.

\section{R-System Thermodynamics}

First, we will focus in the R-system. The surface gravity for this
system is

\begin{equation}
\kappa_{R}=\frac{1}{2}\left(\frac{\partial M}{\partial\mathcal{P}_{R}}\right)_{J,Q}=\frac{1}{2}\left[\left(\frac{\partial M}{\partial\mathcal{P}_{+}}\right)_{\mathcal{P}_{-},J,Q}\left(\frac{\partial\mathcal{P}_{+}}{\partial\mathcal{P}_{R}}\right)_{J,Q}+\left(\frac{\partial M}{\partial\mathcal{P}_{-}}\right)_{\mathcal{P}_{+},J,Q}\left(\frac{\partial\mathcal{P}_{-}}{\partial\mathcal{P}_{R}}\right)_{J,Q}\right]\label{eq:krfirst}\end{equation}

Since $\mathcal{P}_{R}=\mathcal{P}_{+}+\mathcal{P}_{-}$ , we get

\begin{equation}
\frac{\partial\mathcal{P}_{+}}{\partial\mathcal{P}_{R}}+\frac{\partial\mathcal{P}_{-}}{\partial\mathcal{P}_{R}}=1,\label{eq:aux1}\end{equation}
and equation (\ref{eq:diff}) gives

\begin{equation}
\kappa_{+}\left(\frac{\partial\mathcal{P}_{+}}{\partial\mathcal{P}_{R}}\right)_{J,Q}=\kappa_{-}\left(\frac{\partial\mathcal{P}_{-}}{\partial\mathcal{P}_{R}}\right)_{J,Q}.\label{eq:aux2}\end{equation}
Hence, using equations (\ref{eq:aux1}) and (\ref{eq:aux2}) we can
write

\begin{eqnarray}
\left(\frac{\partial\mathcal{P}_{+}}{\partial\mathcal{P}_{R}}\right)_{J,Q} & = & \frac{\kappa_{-}}{\kappa_{+}+\kappa_{-}}\label{eq:aux3}\\
\left(\frac{\partial\mathcal{P}_{-}}{\partial\mathcal{P}_{R}}\right)_{J,Q} & = & \frac{\kappa_{+}}{\kappa_{+}+\kappa_{-}}.\label{eq:aux4}\end{eqnarray}

On the other hand, using equation (\ref{eq:sum}) we obtain

\begin{eqnarray}
\left(\frac{\partial M}{\partial\mathcal{P}_{+}}\right)_{\mathcal{P}_{-},J,Q} & = & \frac{\mathcal{P}_{+}}{l^{2}}-\frac{J^{2}}{4\mathcal{P}_{+}^{3}}-\frac{Q^{2}}{4\mathcal{P}_{+}}=\kappa_{+}\label{eq:aux5}\\
\left(\frac{\partial M}{\partial\mathcal{P}_{-}}\right)_{\mathcal{P}_{+},J,Q} & = & \frac{\mathcal{P}_{-}}{l^{2}}-\frac{J^{2}}{4\mathcal{P}_{-}^{3}}-\frac{Q^{2}}{4\mathcal{P}_{-}}=\kappa_{-}.\label{eq:aux6}\end{eqnarray}

Therefore, putting equations (\ref{eq:aux3}),(\ref{eq:aux4}),(\ref{eq:aux5})
and (\ref{eq:aux6}) into (\ref{eq:krfirst}) we obtain

\begin{equation}
\kappa_{R}=\frac{\kappa_{+}\kappa_{-}}{\kappa_{+}+\kappa_{-}}\end{equation}
or

\begin{equation}
\frac{1}{\kappa_{R}}=\frac{1}{\kappa_{+}}+\frac{1}{\kappa_{-}}.\label{eq:gravityR}\end{equation}
This equation shows that the temperature for the R-system satisfies

\begin{equation}
\frac{1}{T_{R}}=\frac{1}{T_{+}}+\frac{1}{T_{-}},\label{eq:temperatureR}\end{equation}
while the entropy can be written as

\begin{eqnarray}
S_{R} & = & 4\pi\mathcal{P}_{R}=4\pi\left(\mathcal{P}_{+}+\mathcal{P}_{-}\right)=S_{+}+S_{-}.\label{eq:entropyR}\end{eqnarray}

The electric potential is obtained by the expression

\begin{eqnarray}
\Phi_{R} & = & \left(\frac{\partial M}{\partial Q}\right)_{J,\mathcal{P}_{R}}=\left(\frac{\partial M}{\partial\mathcal{P}_{+}}\right)_{J,Q,\mathcal{P}_{-}}\left(\frac{\partial\mathcal{P}_{+}}{\partial Q}\right)_{J,\mathcal{P}_{R}}+\left(\frac{\partial M}{\partial\mathcal{P}_{-}}\right)_{J,Q,\mathcal{P}_{+}}\left(\frac{\partial\mathcal{P}_{-}}{\partial Q}\right)_{J,\mathcal{P}_{R}}+\left(\frac{\partial M}{\partial Q}\right)_{J,\mathcal{P}_{+},\mathcal{P}_{-}}.\label{eq:firfirst}\end{eqnarray}

Since $\mathcal{P}_{R}=\mathcal{P}_{+}+\mathcal{P}_{-}$ , we have

\begin{equation}
\left(\frac{\partial\mathcal{P}_{+}}{\partial Q}\right)_{J,\mathcal{P}_{R}}+\left(\frac{\partial\mathcal{P}_{-}}{\partial Q}\right)_{J,\mathcal{P}_{R}}=0\label{eq:aux7}\end{equation}
and equation (\ref{eq:diff}) gives

\begin{equation}
\left(\frac{\partial\mathcal{P}_{+}}{\partial Q}\right)_{J,\mathcal{P}_{R}}=-\frac{1}{2}\frac{\left(\Phi_{+}-\Phi_{-}\right)}{\left(\kappa_{+}+\kappa_{-}\right)}\label{eq:aux8}\end{equation}

On the other hand, using equation (\ref{eq:sum}) we obtain

\begin{eqnarray}
\left(\frac{\partial M}{\partial Q}\right)_{J,\mathcal{P}_{+},\mathcal{P}_{-}} & = & \frac{1}{2}\left(\Phi_{+}+\Phi_{-}\right).\label{eq:aux11}\end{eqnarray}

Therefore, putting equations (\ref{eq:aux5}),(\ref{eq:aux6}),(\ref{eq:aux7}),(\ref{eq:aux8}),
and (\ref{eq:aux11}) into (\ref{eq:firfirst}), we obtain

\begin{equation}
\Phi_{R}=\frac{\left(\Phi_{+}+\Phi_{-}\right)}{2}-\frac{\kappa_{+}-\kappa_{-}}{\kappa_{+}+\kappa_{-}}\frac{\left(\Phi_{+}-\Phi_{-}\right)}{2}.\label{eq:firsecond}\end{equation}

The angular velocity is given by

\begin{equation}
\Omega_{R}=\left(\frac{\partial M}{\partial J}\right)_{Q,\mathcal{P}_{R}}=\left(\frac{\partial M}{\partial\mathcal{P}_{+}}\right)_{J,Q,\mathcal{P}_{-}}\left(\frac{\partial\mathcal{P}_{+}}{\partial J}\right)_{Q,\mathcal{P}_{R}}+\left(\frac{\partial M}{\partial\mathcal{P}_{-}}\right)_{J,Q,\mathcal{P}_{+}}\left(\frac{\partial\mathcal{P}_{-}}{\partial J}\right)_{Q,\mathcal{P}_{R}}+\left(\frac{\partial M}{\partial J}\right)_{Q,\mathcal{P}_{+},\mathcal{P}_{-}}.\label{eq:auxjr}\end{equation}

Since $\mathcal{P}_{R}=\mathcal{P}_{+}+\mathcal{P}_{-}$ , we have

\begin{equation}
\left(\frac{\partial\mathcal{P}_{+}}{\partial J}\right)_{Q,\mathcal{P}_{R}}+\left(\frac{\partial\mathcal{P}_{-}}{\partial J}\right)_{Q,\mathcal{P}_{R}}=0\label{eq:auxjr1}\end{equation}

and eqution (\ref{eq:diff}) gives

\begin{equation}
\left(\frac{\partial\mathcal{P}_{+}}{\partial J}\right)_{Q,\mathcal{P}_{R}}=-\frac{1}{2}\frac{\left(\Omega_{+}-\Omega_{-}\right)}{\kappa_{+}+\kappa_{-}}\label{eq:auxjr2}\end{equation}

On the other hand, using equation (\ref{eq:sum}) we obtain

\begin{eqnarray}
\left(\frac{\partial M}{\partial J}\right)_{Q,\mathcal{P}_{+},\mathcal{P}_{-}} & = & \frac{1}{2}\left(\Omega_{+}+\Omega_{-}\right)\label{eq:auxjr5}\end{eqnarray}

Thus, putting equations (\ref{eq:aux5}),(\ref{eq:aux6}),(\ref{eq:auxjr1}),(\ref{eq:auxjr2})
and (\ref{eq:auxjr5}) into (\ref{eq:auxjr}), we obtain

\begin{equation}
\Omega_{R}=\frac{\left(\Omega_{+}+\Omega_{-}\right)}{2}-\frac{\kappa_{+}-\kappa_{-}}{\kappa_{+}+\kappa_{-}}\frac{\left(\Omega_{+}-\Omega_{-}\right)}{2}.\label{eq:jr}\end{equation}

Finally, the integral and differential mass formulae for the R-system
are

\begin{eqnarray}
M & = & \kappa_{R}\mathcal{P}_{R}+\frac{1}{2}\Phi_{R}Q+\frac{1}{4}Q^{2}+\Omega_{R}J\\
dM & = & 2\kappa_{R}d\mathcal{P}_{R}+\Phi_{R}dQ+\Omega_{R}dJ,\end{eqnarray}
that corresponds to what is expected as generalization of equations
(\ref{eq:generalIntFirstLaw}) and (\ref{eq:generalDiffFirstLaw}).

\section{L-System Thermodynamics}

Now, we will turn our attention to the L-system. The surface gravity
for this system is

\begin{equation}
\kappa_{L}=\frac{1}{2}\left(\frac{\partial M}{\partial\mathcal{P}_{L}}\right)_{J,Q}=\frac{1}{2}\left[\left(\frac{\partial M}{\partial\mathcal{P}_{+}}\right)_{J,Q,\mathcal{P}_{-}}\left(\frac{\partial\mathcal{P}_{+}}{\partial\mathcal{P}_{L}}\right)_{J,Q}+\left(\frac{\partial M}{\partial\mathcal{P}_{-}}\right)_{J,Q,\mathcal{P}_{+}}\left(\frac{\partial\mathcal{P}_{-}}{\partial\mathcal{P}_{L}}\right)_{J,Q}\right]\label{eq:klfirst}\end{equation}

Since $\mathcal{P}_{L}=\mathcal{P}_{+}-\mathcal{P}_{-}$ , we have

\begin{equation}
\frac{\partial\mathcal{P}_{+}}{\partial\mathcal{P}_{L}}-\frac{\partial\mathcal{P}_{-}}{\partial\mathcal{P}_{L}}=1,\label{eq:aux12}\end{equation}
and using the eqution (\ref{eq:diff}) we obtain

\begin{equation}
\kappa_{+}\frac{\partial\mathcal{P}_{+}}{\partial\mathcal{P}_{L}}=\kappa_{-}\frac{\partial\mathcal{P}_{-}}{\partial\mathcal{P}_{L}}.\label{eq:aux13}\end{equation}
Thus, using equations (\ref{eq:aux12}) and (\ref{eq:aux13}) we can
write

\begin{eqnarray}
\left(\frac{\partial\mathcal{P}_{+}}{\partial\mathcal{P}_{L}}\right)_{J,Q} & = & \frac{\kappa_{-}}{\kappa_{-}-\kappa_{+}}\label{eq:aux14}\\
\left(\frac{\partial\mathcal{P}_{-}}{\partial\mathcal{P}_{L}}\right)_{J,Q} & = & \frac{\kappa_{+}}{\kappa_{-}-\kappa_{+}}.\label{eq:aux15}\end{eqnarray}

On the other hand, equation (\ref{eq:sum}) gives

\begin{eqnarray}
\left(\frac{\partial M}{\partial\mathcal{P}_{+}}\right)_{J,Q,\mathcal{P}_{-}} & = & \frac{\mathcal{P}_{+}}{l^{2}}-\frac{Q^{2}}{4\mathcal{P}_{+}}=\kappa_{+}\label{eq:aux16}\\
\left(\frac{\partial M}{\partial\mathcal{P}_{-}}\right)_{J,Q,\mathcal{P}_{+}} & = & \frac{\mathcal{P}_{-}}{l^{2}}-\frac{Q^{2}}{4\mathcal{P}_{-}}=\kappa_{-}.\label{eq:aux17}\end{eqnarray}

Therfore, putting equations (\ref{eq:aux14}),(\ref{eq:aux15}),(\ref{eq:aux16})
and (\ref{eq:aux17}) into (\ref{eq:klfirst}) we obtain

\begin{equation}
\kappa_{L}=\frac{\kappa_{+}\kappa_{-}}{\kappa_{-}-\kappa_{+}},\end{equation}
or

\begin{equation}
\frac{1}{\kappa_{L}}=\frac{1}{\kappa_{+}}-\frac{1}{\kappa_{-}}.\label{eq:gravityL}\end{equation}
This equation shows that the temperature for the L-system satisfies

\begin{equation}
\frac{1}{T_{L}}=\frac{1}{T_{+}}-\frac{1}{T_{-}},\label{eq:temperatureL}\end{equation}
and the entropy is

\begin{eqnarray}
S_{L} & = & 4\pi\mathcal{P}_{L}=4\pi\left(\mathcal{P}_{+}-\mathcal{P}_{-}\right)=S_{+}-S_{-}.\label{eq:entropyL}\end{eqnarray}

On the other side, the electric potential is give by

\begin{eqnarray}
\Phi_{L} & = & \left(\frac{\partial M}{\partial Q}\right)_{J,\mathcal{P}_{L}}=\left(\frac{\partial M}{\partial\mathcal{P}_{+}}\right)_{J,Q,\mathcal{P}_{-}}\left(\frac{\partial\mathcal{P}_{+}}{\partial Q}\right)_{J,\mathcal{P}_{L}}+\left(\frac{\partial M}{\partial\mathcal{P}_{-}}\right)_{J,Q,\mathcal{P}_{+}}\left(\frac{\partial\mathcal{P}_{-}}{\partial Q}\right)_{J,\mathcal{P}_{L}}+\left(\frac{\partial M}{\partial Q}\right)_{J,\mathcal{P}_{+},\mathcal{P}_{-}},\label{eq:filfirst}\end{eqnarray}

Since $\mathcal{P}_{L}=\mathcal{P}_{+}-\mathcal{P}_{-}$ , we have

\begin{equation}
\left(\frac{\partial\mathcal{P}_{+}}{\partial Q}\right)_{\mathcal{P}_{L}}+\left(\frac{\partial\mathcal{P}_{-}}{\partial Q}\right)_{\mathcal{P}_{L}}=0\label{eq:aux18}\end{equation}
and using the eqution (\ref{eq:diff}) is easily to obtain

\begin{equation}
\left(\frac{\partial\mathcal{P}_{+}}{\partial Q}\right)_{J,\mathcal{P}_{-}}=-\frac{1}{2}\frac{\left(\Phi_{+}-\Phi_{-}\right)}{\kappa_{+}-\kappa_{-}}\label{eq:aux19}\end{equation}

On the other hand, using equation (\ref{eq:sum}) we get

\begin{eqnarray}
\left(\frac{\partial M}{\partial Q}\right)_{J,\mathcal{P}_{+},\mathcal{P}_{-}} & = & \frac{1}{2}\left(\Phi_{+}+\Phi_{-}\right)\label{eq:aux22}\end{eqnarray}

Hence, putting equations (\ref{eq:aux16}),(\ref{eq:aux17}) ,(\ref{eq:aux18}),(\ref{eq:aux19})
and (\ref{eq:aux22}) into (\ref{eq:filfirst}), we have the electric
potential

\begin{equation}
\Phi_{L}=\frac{\left(\Phi_{+}+\Phi_{-}\right)}{2}-\frac{\kappa_{+}+\kappa_{-}}{\kappa_{+}-\kappa_{-}}\frac{\left(\Phi_{+}-\Phi_{-}\right)}{2}.\label{eq:filsecond}\end{equation}

The angular velocity is given by

\begin{equation}
\Omega_{L}=\left(\frac{\partial M}{\partial J}\right)_{Q,\mathcal{P}_{L}}=\left(\frac{\partial M}{\partial\mathcal{P}_{+}}\right)_{J,Q,\mathcal{P}_{-}}\left(\frac{\partial\mathcal{P}_{+}}{\partial J}\right)_{Q,\mathcal{P}_{L}}+\left(\frac{\partial M}{\partial\mathcal{P}_{-}}\right)_{J,Q,\mathcal{P}_{+}}\left(\frac{\partial\mathcal{P}_{-}}{\partial J}\right)_{Q,\mathcal{P}_{L}}+\left(\frac{\partial M}{\partial J}\right)_{Q,\mathcal{P}_{+},\mathcal{P}_{-}}.\label{eq:auxjl}\end{equation}

Since $\mathcal{P}_{L}=\mathcal{P}_{+}-\mathcal{P}_{-}$ , we have

\begin{equation}
\left(\frac{\partial\mathcal{P}_{+}}{\partial J}\right)_{Q,\mathcal{P}_{L}}-\left(\frac{\partial\mathcal{P}_{-}}{\partial J}\right)_{Q,\mathcal{P}_{L}}=0\label{eq:auxjl1}\end{equation}

and eqution (\ref{eq:diff}) gives

\begin{equation}
\left(\frac{\partial\mathcal{P}_{+}}{\partial J}\right)_{Q,\mathcal{P}_{L}}=-\frac{1}{2}\frac{\left(\Omega_{+}-\Omega_{-}\right)}{\kappa_{+}-\kappa_{-}}\label{eq:auxjl2}\end{equation}

On the other hand, using equation (\ref{eq:sum}) we obtain

\begin{eqnarray}
\left(\frac{\partial M}{\partial J}\right)_{Q,\mathcal{P}_{+},\mathcal{P}_{-}} & = & \frac{1}{2}\left(\Omega_{+}+\Omega_{-}\right)\label{eq:auxjl5}\end{eqnarray}

Thus, putting equations (\ref{eq:aux16}),(\ref{eq:aux17}),(\ref{eq:auxjl1}),(\ref{eq:auxjl2})
and (\ref{eq:auxjl5}) into (\ref{eq:auxjl}), we obtain

\begin{equation}
\Omega_{L}=\frac{\left(\Omega_{+}+\Omega_{-}\right)}{2}-\frac{\kappa_{+}+\kappa_{-}}{\kappa_{+}-\kappa_{-}}\frac{\left(\Omega_{+}-\Omega_{-}\right)}{2}\label{eq:jl}\end{equation}

Finally, the integral and differential mass formulae for the L-system
are

\begin{eqnarray}
M & = & \kappa_{L}\mathcal{P}_{L}+\frac{1}{2}\Phi_{L}Q+\frac{1}{4}Q^{2}+\Omega_{L}J\\
dM & = & 2\kappa_{L}d\mathcal{P}_{L}+\Phi_{L}dQ+\Omega_{L}dJ,\end{eqnarray}
that corresponds to equations (\ref{eq:generalIntFirstLaw}) and (\ref{eq:generalDiffFirstLaw})
applied to the L system.

\section{Relationship between the R,L-systems and the BTZ thermodynamics}

The thermodynamic laws of the R, L- systems are related with the BTZ
black hole thermodynamics. Equations (\ref{eq:gravityR}) and (\ref{eq:gravityL})
can be resumed into 

\begin{equation}
\frac{1}{\kappa_{R,L}}=\frac{1}{\kappa_{+}}\pm\frac{1}{\kappa_{-}},\end{equation}
that corresponds exactly with the relation found by Wu\citet{wu2}
for Ker-Newman black hole and A. Larranaga\citet{larr2} for the BTZ
black hole. Since temperature is proportional to surface gravity,
we have a similar expression obtained for equations (\ref{eq:temperatureR})
and (\ref{eq:temperatureL}),

\begin{equation}
\frac{1}{T_{R,L}}=\frac{1}{T_{+}}\pm\frac{1}{T_{-}}.\label{eq:temperatures}\end{equation}

This last relation give us immediately an expression for the Hawking
temperature associated with the BTZ black hole, that corresponds to
the temperature of the outer horizon,

\begin{equation}
T_{H}=T_{+}=\frac{\kappa_{+}}{2\pi},\end{equation}

in terms of the temperatures of the R and L systems. The relation
is 

\begin{equation}
\frac{2}{T_{H}}=\frac{1}{T_{R}}+\frac{1}{T_{L}},\end{equation}

which shows that the Hawking temperature is the harmonic mean of the
R and L temperatures, just as happen in effective string theory or
D-brane physics. \\
Now, lets play some attention to the electric potential and the angular
velocity. Note that using equations (\ref{eq:gravityR}) and (\ref{eq:gravityL})
we can rewrite the R and L electric potentials given by (\ref{eq:firsecond})
and (\ref{eq:filsecond}), as

\begin{equation}
\Phi_{R,L}=\frac{\left(\Phi_{+}+\Phi_{-}\right)}{2}+\frac{\kappa_{R,L}}{\kappa_{L,R}}\frac{\left(\Phi_{+}-\Phi_{-}\right)}{2}.\label{eq:ficomplete}\end{equation}

In a similar way, the angular velocity in equations (\ref{eq:jr})
and (\ref{eq:jl}), can be written as

\begin{equation}
\Omega_{R,L}=\frac{\left(\Omega_{+}+\Omega_{-}\right)}{2}+\frac{\kappa_{R,L}}{\kappa_{L,R}}\frac{\left(\Omega_{+}-\Omega_{-}\right)}{2}.\label{eq:jcomplete}\end{equation}

The form of these equations are exactly the same found for the Kerr-Newman
black hole \citet{wu2}, showing again that the effective string theory
thermodynamics seems to be a universal picture holding also in 2+1
gravity.

\section{conclusion}

In this paper we have shown that the thermodynamics of the (2+1) dimensional
rotating charged BTZ black hole can be constructed from two independient
thermodynamical systems that resemble the right and left modes of
string theory. If one assume that the effective strings ha0ve the
same mass and electric charge that the charged BTZ black hole, there
is a correspondence between the R and L modes thermodynamics and the
thermodynamics of the horizons. 

We have show that the Hawking temperature associated with the black
hole is obtained as the harmonic mean of the temperatures associated
with the R and L systems, just as in the case of stringy thermodynamics.
Moreover, equations (\ref{eq:ficomplete}) and (\ref{eq:jcomplete})
show that the electric potential and the angular velocity of the R
and L systems are related with the electric potential and angular
velocities of the inner and outer horizons with the same relation
obtained by Wu \citet{wu2} for the Kerr-Newman black hole. \\

All these facts suggest that there is a deep connection between string
theory and D-branes with black holes physics that seems to hold in
many cases, not only in General Relativity but also in 2+1 gravity.
Therefore, it is really interesting to investigate if this relation
can give some clue for the understanding of the origin of black hole
entropy.

\end{document}